\documentclass[aps,pre,preprint,superscriptaddress,nofootinbib]{revtex4-1}
\usepackage{graphicx}
\usepackage{amsmath}
\usepackage{amsfonts}
\usepackage{hyperref}
\begin{document}

\title{Spatial distribution of entanglements in thin free-standing films}

\author{Daniel M. Sussman}
\email[]{dsussman@sas.upenn.edu}
\affiliation{Department of Physics and Astronomy, University of Pennsylvania, 209 South 33rd Street, Philadelphia, Pennsylvania 19104, USA}

\begin{abstract}
We simulate entangled linear polymers in free-standing thin film geometries where the confining dimension is on the same scale or smaller than the bulk chain dimensions. We compare both film-averaged and  layer-resolved, spatially  inhomogeneous measures of the polymer structure and entanglement network with theoretical models. We find that these properties are controlled by the ratio of both chain- and entanglement-strand length scales to the film thickness. While the film-averaged entanglement properties can be accurately predicted, we identify outstanding challenges in understanding the spatially resolved character of the heterogeneities in the entanglement network, particularly when the scale of both the entanglement strand and the chain end-to-end vector is comparable to or smaller than the film thickness.
\end{abstract}

\maketitle

\section{Introduction}

The properties of polymer melts under strong confinement conditions has been a subject of intense interest, where much of this attention has focused on elucidating changes in the glassy properties of nanometrically thin samples \cite{Keddie1994}. It is known, though, that strong confinement also affects the properties of thin polymer films in the melt state. For instance, bubble-inflation experiments have revealed a stiffening in the creep compliance of thin films in the rubbery plateau regime \cite{McKenna2005}, a feature which is robust to performing the experiments well above the glass transition temperatures of the films \cite{McKenna2006}. Measurements in nanometrically confined thin films have reported lower viscosity \cite{Masson2002, Rowland2008} and greater mobility \cite{Shin2007}, and the diffusion rate in, e.g., cylindrical nanopores is known to be affected by the relative degree of chain confinement \cite{Saalwachter2015}. Recent work on polymer nanocomposites has revealed that the (local) confinement effects introduced by nanoparticles can lead to interesting chain conformation and entanglement heterogeneities in a sample\cite{Clarke2012,Liu2013,Weir2016}, which can in turn affect the global system properties.

An emerging theme is the difficulty of understanding entangled polymer melts when the dimensions of the confining volume are comparable to or smaller than the typical end-to-end distance of the chains, $R_{ee}$. This difficulty is compounded in the ``ultra-thin'' films typically simulated, where the confining dimensions may be small or comparable not only to the chain conformational scale but also to the typical length scale associated with the entanglement strands, $L_e$. While techniques to predict chain-scale conformational changes are relatively well-developed \cite{Cavallo2005b,Sussman2014}, the connections between these chain or sub-chain structural changes and system dynamics are much more poorly understood. 

Experiments on entangled thin films have interpreted changes in macroscopic system properties in terms of an increase in the entanglement molecular weight, $M_e$, and a decrease in interchain entanglement density in response to chain-scale confinement \cite{Si2005,Liu2015}. Consistent with this observation, coarse-grained simulations analyzed using ``primitive path'' (PP)-determining methods directly observed a change in the number of interchain entanglements under both geometric \cite{Sussman2014, Cavallo2005, Meyer2007, Barrat2007} and phase-induced confinement \cite{Ganesan2015} conditions. Again, though, the connections between structural entanglement statistics -- such as the number of entanglements per chain ($\langle Z\rangle$), the tube diameter ($d_T$), and the number of monomers per entanglement ($N_e$) -- and the dynamical properties of the system are unclear. The classic Doi-Edwards model provides a natural framework for addressing this question in bulk, homogeneous systems \cite{DoiEdwards}, but some modifications to the theory are needed under conditions of strong confinement. For instance, recent simulations of polymers confined in nanopores suggest that the changes in chain diffusion are qualitatively consistent with stronger confinement leading to lower $\langle Z\rangle$, but that a straightforward reptation-model estimate of diffusivity seems quantitatively inaccurate \cite{Tung2015}.

We take one possible interpretation of these findings to be that the primitive path measures of entanglement in simulations indeed directly inform system dynamics, but that a model that explicitly treats any spatial inhomogeneity of entanglement statistics must be used to accurately predict dynamical properties. This challenge is two-fold, since at the moment there are few theories that might be used to predict the inhomogeneities of any entanglement statistics in the first place, let alone theories that also connect those inhomogeneities to other measurable single-chain or macroscopic observables of the system. One appealing model of entanglements near a free surface simply posits that at the surface there are only half as many chains with which to entangle. If there are $\langle Z\rangle$ inter-chain entanglements per chain for an equivalent bulk system, then, at the surface there might be $\langle Z\rangle/2$ entanglements, with some smooth, monotonic interpolation in between \cite{Russell1996,Si2005}. At the same time, it is known that the chain conformations are perturbed up to a distance $\sim R_{ee}$ from a boundary, and a recent theory has been proposed connecting the orientational correlations that arise from such conformational perturbations to entanglement statistics \cite{Sussman2012,Sussman2013}. Such correlations could be relevant for not only confined chain systems, but also bulk chains systems undergoing deformations, where it is known that chain-scale orientation occurs together with a decrease in the effective entanglement number \cite{Baig2010,Khomami2014}.

In this paper we perform simulations with the goal of collecting statistics on the average and spatially resolved distribution of both chain conformational properties and entanglement statistics of confined polymer chains. We hope that this data both provides a robust test of the above theories and encourages the development of other theoretical models of chain entanglement in confined and non-equilibrium polymer systems. We focus on preparing equilibrated free-standing thin films, using a standard bead-spring coarse graining, equilibration protocol, and PP-extraction method, described in more detail below. We have chosen to work with monodisperse polymers composed of $N=2000$ monomers, in films of thickness $\sim 13,$ $20,$ $35,$ $48$, and $58$ in units of the monomer diameter. We also present results on shorter-chain films derived from these base data sets. These parameters were chosen to probe regimes in which the film thickness is comparable to or larger than the bulk tube diameter, while being simultaneously comparable to or smaller than characteristic end-to-end chain scale.

The remainder of the paper is organized as follows. Section \ref{sec:methods} describes in detail the simulation model used and the protocol followed to equilibrate the systems studied in this work. Section \ref{sec:results} presents the numerical results for the film-averaged and the spatially resolved chain entanglement statistics found in our free-standing films, as well as spatially resolved orientational correlations on both the chain end-to-end and PP length scales. We close in Section \ref{sec:disc} with a discussion of our results.

\section{Methods}\label{sec:methods}

We perform molecular dynamics simulations of a commonly used coarse-grained model polymer \cite{Grest1990} prepared in thin film geometries. Our systems have $N_T=5\times 10^5$ total particles of diameter $\sigma$, composed of $M=250$ monodisperse chains of length $N=2000$. The non-bonded interactions between particles $i$ and $j$ are specified by 
\begin{equation}
V_{ij}^{nb}=4\epsilon \left( \left( \frac{\sigma}{r_{ij}} \right)^{12} - \left( \frac{\sigma}{r_{ij}} \right)^6 \right) -4\epsilon \left( \left( \frac{\sigma}{r_{c}} \right)^{12} - \left( \frac{\sigma}{r_{c}} \right)^6 \right)
\end{equation}
for $r_{ij}<r_c$ and $V_{ij}=0$ for $r_{ij}>r_c$. Here $\epsilon$ sets the energy scale, and we take the range of the non-bonded interactions to be $r_c=2.5$. For the bonded interactions we use a very stiff harmonic  potential,
\begin{equation}
V_{ij}^{b}=\frac{k}{2}\left(r_{ij}-\sigma\right)^2,
\end{equation}
where $k=2000\epsilon/\sigma^2$. In this paper we report results in reduced LJ simulation units, e.g., the temperature $T=kT^*/\epsilon$ and the time $\tau=\tau^*\sqrt{\epsilon/m\sigma^2}$, where $T^*$ and $t^*$ are defined in laboratory units and $m$ is the mass of a particle. All of our simulations were run in the NVT ensemble at $T=1.0$ and a timestep of $\delta t = 0.0025\tau$ using the HOOMD-blue simulation package \cite{hoomd-blue0,hoomd-blue} .

Five base systems are considered in this work, corresponding to different effective film thicknesses, $h_{eff}$. The films were prepared using a common protocol, which we document here for completeness, given that for these large systems the overwhelming bulk of processing time is spent preparing independent equilibrated samples. We begin by pre-packing non-interacting random walks with the correct single-chain statistics \cite{Auhl2003} in a simulation cell with the dimensions of the desired film. By treating two of the three dimensions as periodic and the other ($z$-) dimension as a reflecting wall at this step we are able to start with a configuration of chains that will not be greatly perturbed when the $z$ direction of the box is expanded to expose the surfaces of the film to vacuum. We then use a collection of chain-altering Monte Carlo (MC) moves that simultaneously reduce the density fluctuations of the pre-packed chains\cite{Auhl2003} and respect the single-chain statistics imposed by the model potentials and by the thin film geometry \cite{SussmanBitbucket}. From these non-interacting chain configurations the box was expanded in the $z$ direction to allow for a free-standing film, the LJ and bonded interactions were slowly turned on, and molecular dynamics using the potentials and parameters above were begun. The film thickness, $h_{eff}$, is determined by fitting the density profile of the films in the $z$ direction, $\rho(z)$, to
\begin{equation}
\rho(z) =\frac{c_1}{2}\left[1+\tanh \left( \frac{|z|-\xi}{1/\lambda}\right) \right].
\end{equation}
This mean-field-like functional form fits the density profiles very well; here $c_1$ is determined by the overall density and the simulation box size, $\xi$ is a measure of film thickness, and $\lambda$ describes the extent of density fluctuations near the surface. These density fluctuations combine the effect of equilibrium fluctuations whose scale depends on the bulk correlation length and surface capillary waves whose contribution to the interfacial width would grow logarithmically with the size of the simulation box \cite{Buff1965,Hapke1998}. For our films at $T=1.0$ the interfacial width is small, and we find $\lambda \approx 1$ for all of our films. We take a convenient measure of the effective film thickness to be $h_{eff}=2(\xi+\lambda)$, a spatial extent which for our films encompasses at least $99\%$ of the particles in the simulation box. By this measure, the free-standing films simulated in this work have thicknesses $h_{eff}/\sigma =$ $13.3,$ $20,$ $35,$ $48$, and $58$.

We are interested in the spatial distribution of entanglements in these simulated films, and we use the Z1 algorithm to extract the entanglement statistics of our polymer melts. For bulk systems the two geometric methods of extracting primitive paths (Z1 \cite{Kroger2005,Kroger2009,Kroger2009b,Kroger2007} and CReTA \cite{Creta}) are expected to give similar global entanglement measures, and for confined thin films this expectation has been explicitly verified \cite{Sussman2014}.  We note that the primitive paths revealed by the Z1 algorithm vary slightly from the ideal random-walk statistics envisioned by the conventional tube model, with non-Gaussian distributions of primitive path segment sizes and weak angular correlations between consecutive primitive path steps. However, at the level of theoretical comparisons that will be made in this paper we do not expect the non-Gaussianities to be important.

To speed up equilibration we implemented double-rebridging, connectivity altering moves \cite{Auhl2003, Karayiannis2002} on graphics processing units in the HOOMD-blue framework \cite{SussmanBitbucket}. Modeled after the protocol outlined in Ref. \cite{Auhl2003}, we check roughly half of the particles for chain-length-preserving bond swaps every $0.125\tau$; this longer time leads to fewer proposed reverse swaps compared to checking every few time steps, and reduces the impact of the bond-swapping algorithm on overall simulation performance. We note that we have expanded the algorithm to be able to handle not only bonds and angular potentials, but also dihedral potentials; results generated by treating chains with dihedral potentials will be discussed in a future publication.

As an initial consideration for system equilibration, Auhl et al. noted that one needs $\mathcal{O}(10)$ pivoting moves to decorrelate the end-to-end vector of an isolated chain  \cite{Auhl2003}. In the context of a dense melt, bond-swapping alone decorrelates subchains on which at least $\mathcal{O}(10)$ bond-swapping moves have been performed. To equilibrate the chains on all length scales, then, the molecular dynamics must be run up to the decorrelation time of the longest subchains that have not participated in $\mathcal{O}(10)$ connectivity-altering Monte Carlo moves. Another consideration is that individual monomers should be able to sample bond swaps with a suitable number of exchange partners. For these linear chains the typical distance between monomers of the same index is $\sim (\rho/(N/2))^{-1/3}$; if the monomers were to diffuse with Rouse-like dynamics the typical time to diffuse this distance, $t_{s}$, corresponds  to the Rouse time of a chain of length $N_{eff}=(N/2)^{2/3}$ \cite{Auhl2003}. For our chains of length $N=2000$ this corresponds to diffusive time scales of order $t_s \sim 10^4\tau$. This use of Rouse-like dynamics to estimate the relevant time scale arises because the Monte Carlo bond-swapping procedure allows chain reconfigurations that remove previously existing entanglement constraints. This effect is seen in the monomer mean-square displacements, which under the action of the bond-swapping moves cross over directly from a $1/2$ power law to diffusive dynamics -- as in the Rouse model -- rather than crossing over from a $1/2$ power law to a more slowly growing function of time as would be expected if the system were evolving under dynamics that preserved chain connectivity. In practice this is a conservative estimate of the typical time to diffuse the distance between monomers of the same index.
 
\begin{figure}
\centerline{
\includegraphics[width=3.1in]{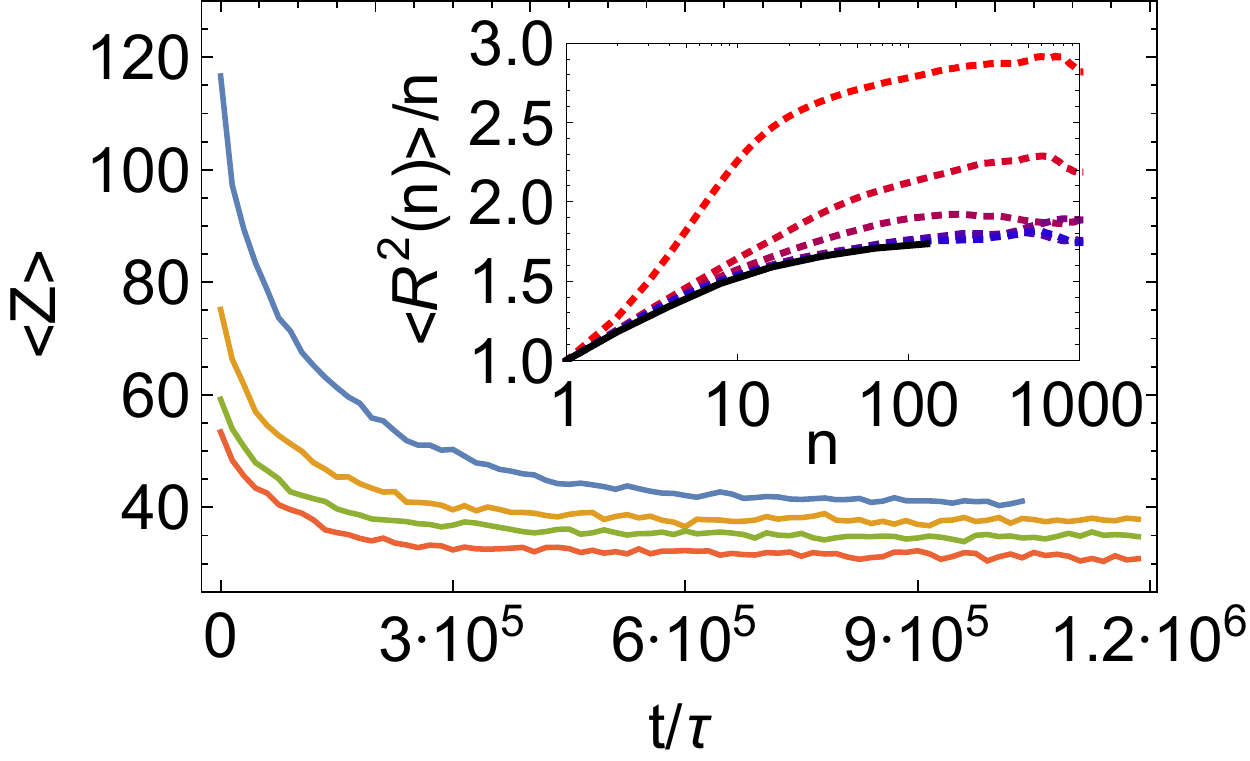}
}
\caption{\label{fig:equil} Average number of kinks per chain measured by the Z1 algorithm during the equilibration of systems with $M=250$ chains each of length $N=2000$. From top to bottom the curves correspond to a bulk system followed by thin films of size $h_{eff}=35\sigma$, $h_{eff}=20\sigma$, and $h_{eff}=13.3\sigma$. Inset. The mean-square internal distance along a chain \cite{Auhl2003} as a function of monomer index separation along a chain. Shown are curves for the equilibration of a bulk system at (top to bottom, dotted curves) $t/\tau = 0,\ 1.75\times 10^5,\ 3.5\times 10^5,\ 7.0\times 10^5,\ 8.75\times 10^5,\ 1.05\times 10^6$. The solid curve is the ``target function'' from Auhl et al.\cite{Auhl2003}}
\end{figure}

Thus, we first equilibrate our system systems by running molecular dynamics with the MC bond swaps for at least $t_{eq}=7.5\times10^5\tau$, a time long enough that every monomer is involved in  $\mathcal{O}(10)$ bond-swapping moves and long compared to the diffusive time for a monomer to travel the typical distance separating monomers of the same index. Since it is impractical to simulate the system for long enough that a typical monomer could diffuse more than the end-to-end distance of a chain, we check equilibration by (1) looking at the distribution of single-chain conformational statistics \cite{Auhl2003} and (2) using the Z1 algorithm to look at the entanglement statistics of our systems \cite{Hoy2005}. Figure \ref{fig:equil} shows illustrative examples  of the time evolution of these measures during the equilibration phase. We note that the conformational comparison in Fig. \ref{fig:equil} is to polymers using finite extensible non-linear elastic potentials,
\begin{equation}
V_{ij,FENE}^b = \left\{  \begin{array}{cc} -0.5k R_0^2 \log \left(1-(r_{ij}/R_0)^2 \right)& r_{ij} \leq R_0 \\
\infty & r_{ij} > R_0
\end{array}
 \right.,
\end{equation}
with parameters $k=30\epsilon/\sigma^2$ and $R_0=1.5\sigma$, rather than the stiff harmonic springs used in this work. Additionally, to treat cohesive thin films we use a LJ potential with both repulsive and attractive regions, whereas Auhl et al. use a purely repulsive LJ potential for the non-bonded interactions.. These differences lead to a slight change the expected bond length $\langle b^2\rangle^{1/2}$ for the single chains, here $\langle b^2\rangle^{1/2}=0.9987\sigma$ whereas the system of  Auhl et al. has $\langle b^2\rangle^{1/2}=0.97\sigma$ at the same density and temperature \cite{Auhl2003}. The angular statistics are identical, though, and on the scale of the plot the effect of this difference for the mean-square internal distance statistics is negligible. 

For the entanglement statistics, in our equilibrium bulk $N=2000$ systems of monomer density $\rho_s = 0.86$ we find that the primitive path segment size, $a_{pp} \equiv \langle R_{ee}^2\rangle /L_{pp}$ where $L_{pp}$ is the primitive path contour length, is $a_{pp}/\sigma=12.5 \pm 0.4$, the mean number of kinks per chain is $\langle Z \rangle  =40.8 \pm 0.20$, and $N_e = 49.1 \pm 0.28$. These numbers are consistent with values reported in the literature for this model polymer \cite{Hoy2005}. In the remainder of the paper we will take the PP segment size as our characteristic scale of the entanglement strands, $L_e\equiv a_{pp,bulk}$; again, the slight non-Gaussianities of the primitive paths imply that these length scales are, in principle, close but distinct quantities.

Once the films are equilibrated as described above, we continue to run the molecular dynamics with connectivity-altering MC moves. We record and analyze our system every $5000\tau$, which corresponds to each chain experiencing $\approx 100$ successful bond swaps. By the criteria described above our $N=2000$ chains are still correlated: on average subchains of length $\mathcal{O}(500)$ and smaller have yet to experience $\mathcal{O}(10)$ bond swaps, and since $5000\tau$ is much less than the Rouse time of $N=500$ chains the molecular dynamics does not equilibrate these sub-chains. Additionally, individual monomers have not diffused enough to sample a reasonable number of swap partners. By these criteria a better choice would be to perform our analysis on frames separated by at least $4\times10^4\tau$. This is greater than $t_{s}$, and on this time scale bond-swapping would decorrelate the subchains up to length $\mathcal{O}(120)$, for which the Rouse time is only $\sim 2\times 10^4$. Even less frequent sampling would further decrease the correlations between the samples. However, the primitive path network found in the simulations is very sensitive to both the global chain features (that is, a large change in the positions of the chain ends can affect entanglements even in the middle of that chain) and even a small number of connectivity altering swaps along subchains of size $N_e$. Thus, we find that the primitive paths are decorrelated even on the shorter time scale at which we save our configurations. Furthermore, by separately averaging our results over all saved frames and over only a subset, we have checked that our entanglement results are qualitatively insensitive to the fact that we are not sampling truly independent underlying configurations.

\section{Results}\label{sec:results}

\subsection{Global entanglement statistics}
\begin{figure}
\centerline{
\includegraphics[width=3.1in]{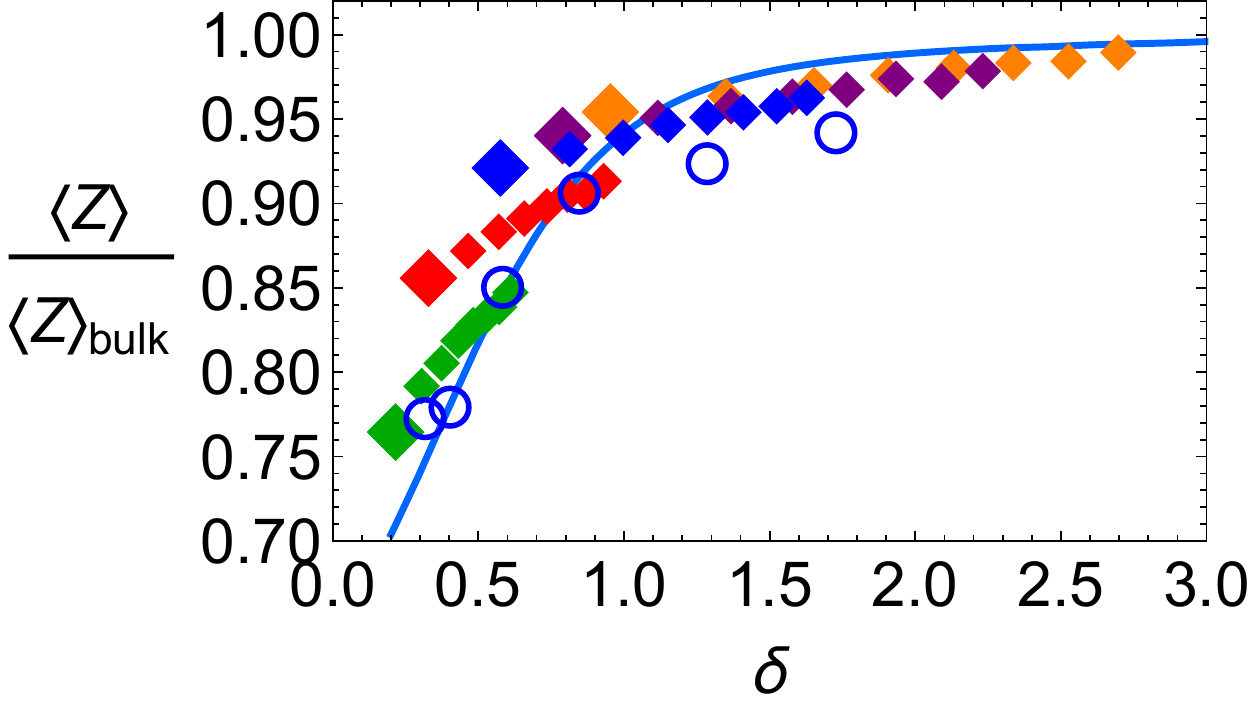}
}
\caption{\label{fig:Ztot} System-averaged reduction of entanglements as a function of confinement. The large diamonds correspond to free-standing films of $N=2000$ chains. Small diamonds correspond to films generated by taking equilibrium configurations of the $N=2000$ chains and subdividing each chain into subchains of length $N=1000,\ 667,\ 500,\ 400,\ 333,\ 285,\ 250$. The open circles are the systems of $N=350$ chains confined between hard amorphous walls from Ref. \cite{Sussman2014}. The solid curve is the theoretical prediction of Ref. \cite{Sussman2014} }
\end{figure}

With the film configurations described above in hand, we first report the film-averaged entanglement statistics of our systems. Using the Z1 algorithm, we find that our free-standing thin films indeed show a smaller number of entanglements per chain, $\langle Z \rangle$, than our bulk system. Figure \ref{fig:Ztot} shows this reduction as a function of the film confinement parameter $\delta \equiv h_{eff}/R_{ee,bulk}$, the effective film thickness divided by the bulk end-to-end chain distance, that has previously been used to collapse similar data \cite{Si2005,Sussman2014}. The figure also shows a theoretical prediction (described below) and simulation data from Ref. \citenum{Sussman2014}. 

To increase the number of data points we can investigate in this representation, we note that we can subdivide the chains in our simulations of length $N=2000$ polymers into subchains of shorter length. For instance, severing the bonds in the center of each chain creates a system of $M=500$ chains of length $N=1000$. Since the $N=2000$ chains have correct intermediate internal separation single-chain statistics, this derived sub-chain system will be in a nearly equilibrium configuration (excluding chain-end effects). We numerical test this assumption of equilibration in the same way as for the full chain systems, i.e., by comparing single-chain conformational statistics and by looking at the number of entanglements per chain detected by the Z1 algorithm. We have confirmed that both the mean-square internal distance statistics of single chains and the entanglement statistics obtained in this way are indeed representative of the entanglement statistics shorter-chain systems by independently equilibrating films of thickness $h_{eff}=35\sigma$ and $h_{eff}=20\sigma$ with the same number of total monomers ($N_T=5\times 10^5$) grouped in chains of length $N=500$. We confirm that the reduction in entanglement density is the same as that obtained by running the Z1 algorithm on systems created by the subdivision method. For the many other subdivided chain systems considered below, we do not repeat the process of independently equilibrating films of the same chain length; we confirm that the single-chain conformational statistics fall on the equilibrium master curve, and take the entanlement data from running the Z1 algorithm on the sub-chain systems to represent the equilibrated value.

Using the subdivision method allows us to infer the entanglement statistics in films of the same thickness but for chains whose length is (approximately) an integer divisor of $N=2000$. Although the lack of collapse of the $N=2000$ chains simulated here with the data from Ref. \cite{Sussman2014} could be potentially attributed to the difference between freestanding and capped thin films, it is clear that the entanglement statistics of these subdivided chains do \emph{not} collapse cleanly when the film thickness is reported in units of $\delta$. Our independent simulations of $N=500$ chains confirms that this lack of collapse cannot be attributed to an artifact arising from studying subdivided-chain systems derived from equilibrated films of longer chains.

One theoretical expectation for collapse of thin film entanglement data with $\delta$, and the source of the theoretical prediction in Fig. \ref{fig:Ztot}, arises from a statistical mechanical framework originally developed for fluids of entangled needles \cite{Szamel1993}. The extension of this theory to flexible chains argues that orientational ordering on the chain scale could lead to a larger tube diameter and hence a reduction in the number of entanglements per chain \cite{Sussman2012,Sussman2013,Sussman2011}. That theory assumes that orientational correlations present on the chain end-to-end scale were directly communicated to the PP scale, and that PP orientational correlations, averaged over the entire system, affected the tube diameter as
\begin{equation}\label{eq:theory}
\frac{d_T}{d_{T,bulk}}\propto \left( \int d\vec{u}_1 d\vec{u}_2 g(\vec{u}_1) g(\vec{u}_2)\sqrt{1-\left(\vec{u}_1\cdot \vec{u}_2\right)^2}\right)^{-1},
\end{equation}
where $g(\vec{u}_i)$ is the orientational distribution of PP $i$. We note that the non-Gaussianities of the primitive paths revealed in simulations were shown to be reasonably well captured by the theory that leads to Eq. ref{eq:theory} \cite{Sussman2012}. Since it is possible to theoretically predict the conformational properties of an ideal melt of chains in the presence of hard, reflecting walls \cite{Cavallo2005b,Sussman2014}, Eq. \ref{eq:theory} can be used to predict from first principles the change in the tube diameter as a function of confinement. By assuming standard Doi-Edwards relations among the properties of the tube, one obtains
\begin{equation}\label{eq:theory2}
\frac{\langle Z(\delta) \rangle}{\langle Z \rangle_{bulk}} = \frac{\langle N_e  \rangle_{bulk}}{\langle N_e(\delta)  \rangle} = \frac{\langle d_T^2  \rangle_{bulk}}{\langle d_T^2(\delta)  \rangle },
\end{equation}
where $d_T/d_{T,bulk}$ is given by Eq. \ref{eq:theory}. As seen in Fig. \ref{fig:Ztot}, this theory was found to be a fairly good model for the film-averaged changes in $\langle Z \rangle$ as a function of film thickness for systems of $N=350$ chains confined by hard, amorphous walls.

In contrast, the theory systematically over-estimates the changes in entanglement for the $N=2000$ free-standing films simulated here. On the one hand, this over-estimate is consistent with the expectation that free-standing films likely induce less chain-scale orientational correlations than hard, reflecting walls (although it is not consistent with the expectation that the film dimensions we consider might be expected to have larger changes to the distribution of chain end-to-end vectors \cite{Cavallo2005b} than are considered in the simplified conformational model of Ref. \cite{Sussman2014}). On the other hand, more concerning is the apparent lack of collapse when the film sizes are reported in units of $\delta$ for our independent simulations of $N=500$ chains and subchain-derived data sets for chains of length $N=2000,$ $1000,$ $666,$ $500,$ $400,$ $333,$ $285,$ and $250$. We emphasize, though, that for the range of film thicknesses studied here (and throughout much of the simulation literature) there is another crucial length scale contributing to the problem: for such thin films the tube diameter itself may be comparable to the film thickness, and the ratio $h_{eff}/L_e$ should also enter any attempted scaling collapse of the entanglement data.

\begin{figure}
\centerline{
\includegraphics[width=3.1in]{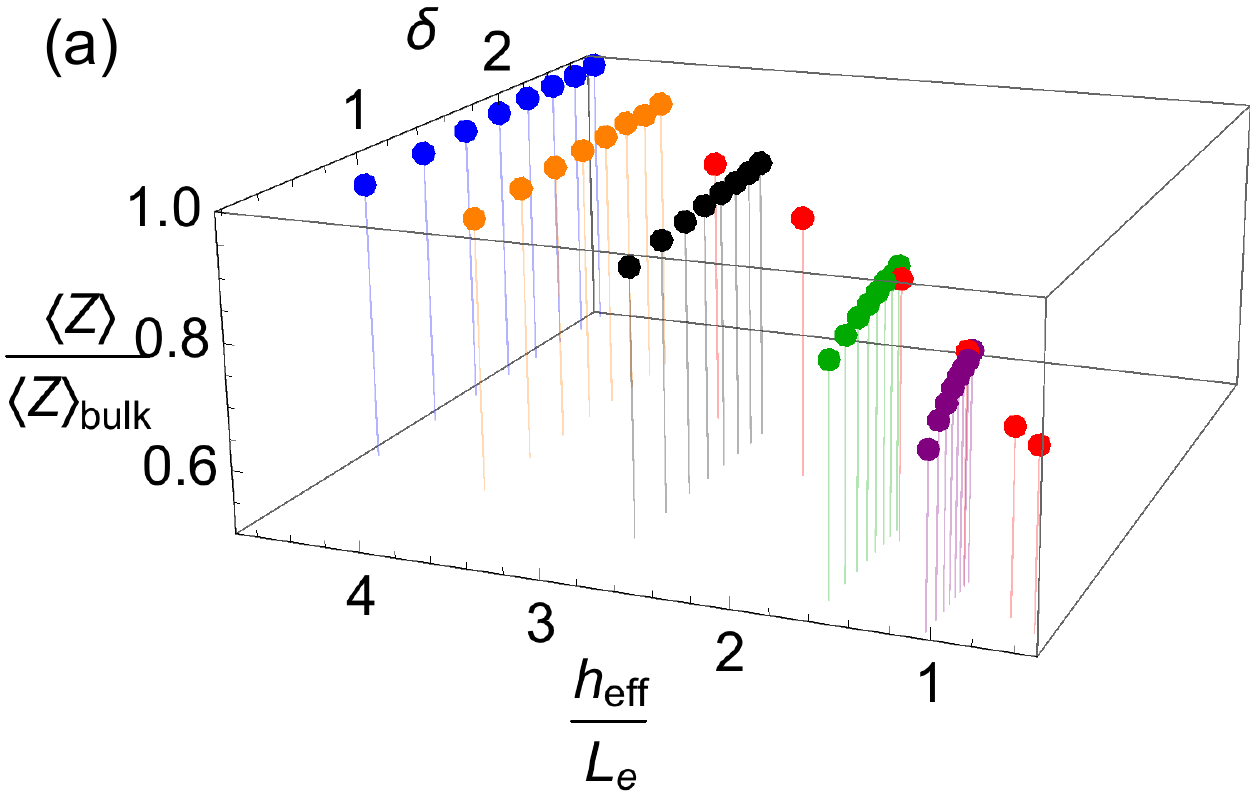}}
\centerline{
\includegraphics[width=3.1in]{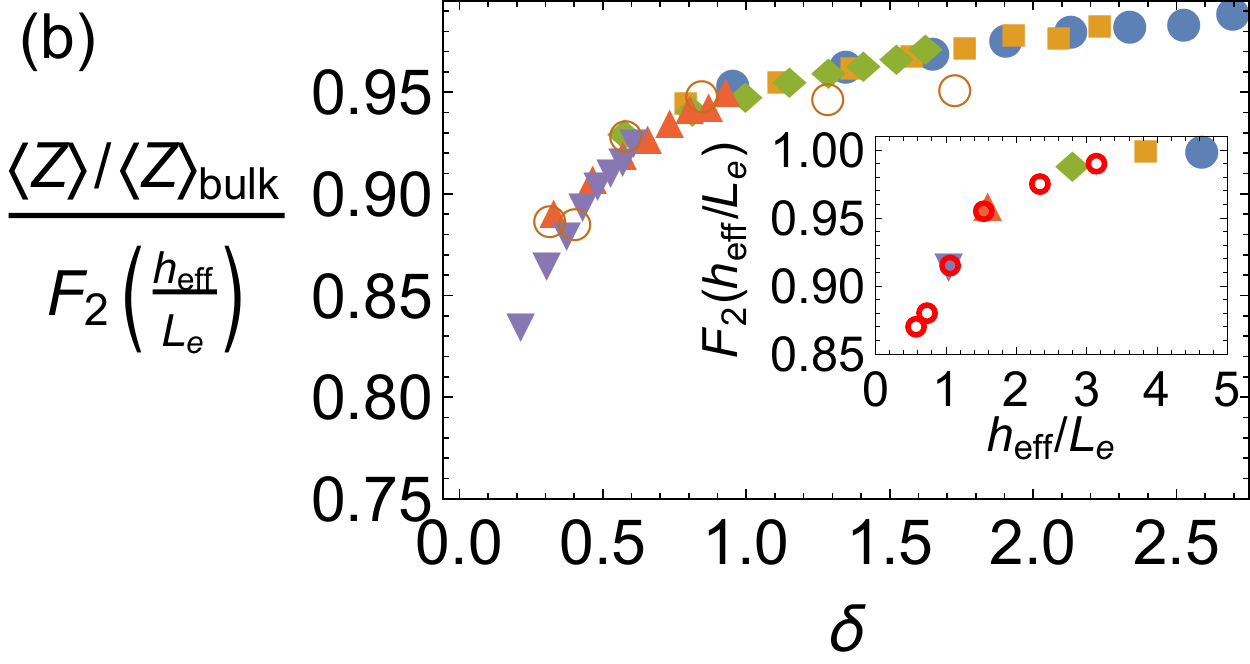}
}
\caption{\label{fig:threeplot} (a) System-averaged reduction of entanglements as a function of the confinement scale relative to both the chain dimension and $L_e$. Data sets of constant $h_{eff}/L_e$ (blue, orange, black, green, and purple points) correspond to the simulations of this work, and the remaining data (red points) correspond to the systems of $N=350$ chains confined between hard amorphous walls from Ref. \cite{Sussman2014}. (b) System-averaged reduction of entanglements normalized by an appoximate function of the dependence on $h_{eff}/L_e$, as described in the text. Inset. The value of the function $F_2(h_{eff}/L_e)$ used to collapse the data sets in the main frame. Identical symbols in the main frame and inset correspond to the same data sets.
}
\end{figure}

Although our data are quite limited, in Fig. \ref{fig:threeplot}a we see that the data for entanglement loss plausibly fall on a single surface when plotted simultaneously as a function of $\delta$ and $h_{eff}/L_e$, a reflection that in very thin films both length scales are of importance in determining the system-averaged reduction in the number of entanglements per chain. We propose an ansatz in which the dependence is separable:
\begin{equation}\label{eq:ansatz}
\frac{\langle Z\rangle}{\langle Z\rangle_{bulk}} = F\left( \delta,h_{eff}/L_e\right) \approx  F_1\left( \delta\right)  F_2\left( h_{eff}/L_e \right).
\end{equation}
To test this, we approximate $F_2\left( h_{eff}/L_e \right)$ by the value needed to collapse the data simulated in this work as a function of $\delta$. As shown in Fig. \ref{fig:threeplot}b, smoothly extrapolating this function also approximately collapses the data for capped films of $N=350$ chains simulated in Ref. \cite{Sussman2014}. The values of $F_2\left( h_{eff}/L_e \right)$ needed to collapse the data are shown in the inset of Fig. \ref{fig:threeplot}b. Thus, we view the ansatz of Eq. \ref{eq:ansatz} as plausible in light of the data available in this work, but caution that this collapse is over a relatively limited range of parameter space. This encourages a future systematic study of entanglement reduction as a function of $\delta$ and $h_{eff}/L_e$. We note that the experimental data from Ref. \cite{Si2005} are all at reasonably large values of $h_{eff}/L_e$, suggesting a possible explanation for the better data collapse observed in that work.

\subsection{Spatial distribution of entanglement and conformational statistics}

\begin{figure}
\centerline{
\includegraphics[width=3.1in]{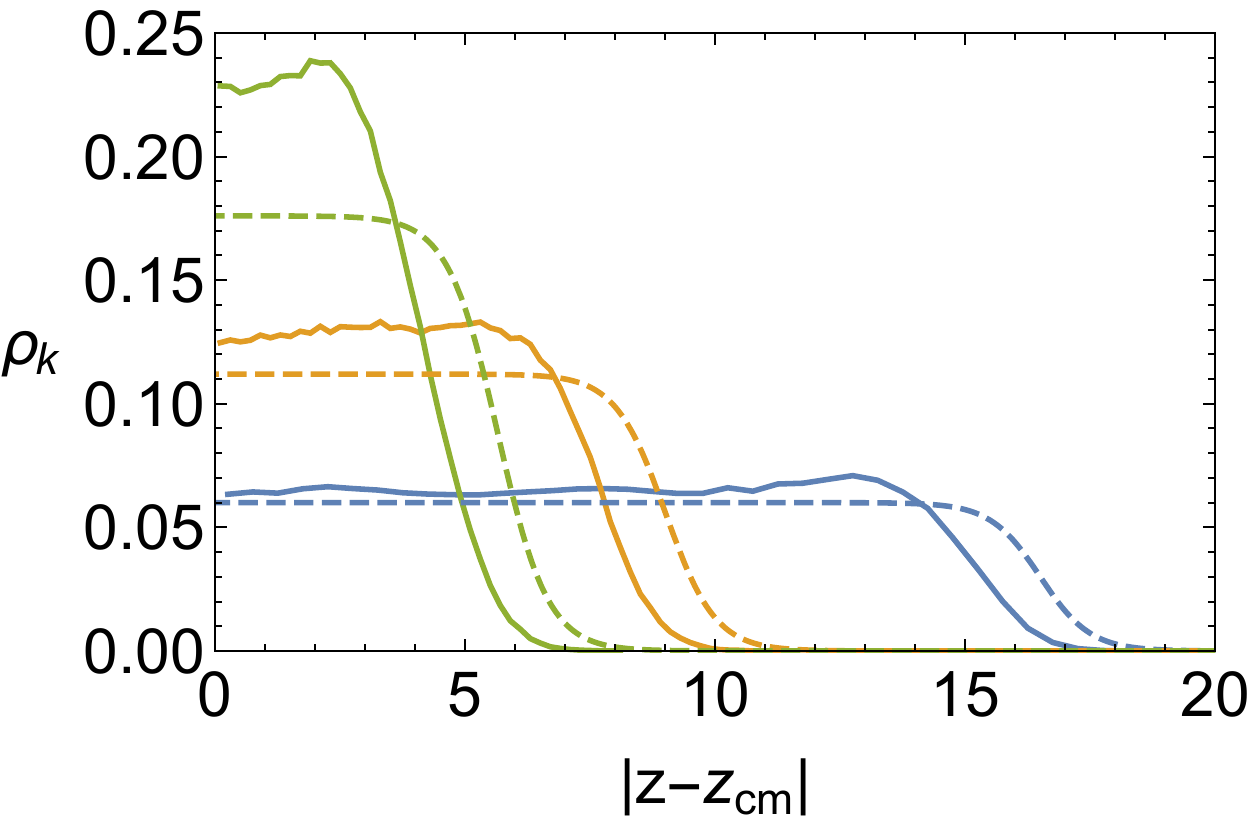}
}
\caption{\label{fig:khist} Probability distribution of entanglements as a function distance in the $z$-direction from plane corresponding to the center of mass of the film (solid curves) and the total density profile of the films (dashed curves). From bottom to top the curves correspond to $N=2000$ chains in free-standing films with effective thickness $h_{eff}=35\sigma$, $h_{eff}=20\sigma$, and $h_{eff}=13.3\sigma$. 
}
\end{figure}

We now focus on the spatial inhomogeneities in the distribution of entanglement and conformational statistics throughout the films. This allows for tests of the spatially resolved properties of the entanglement network that (a) were predicted by the theory described above but (b) were not previously tested due to limits on the amount of data available \cite{Sussman2014}. In this respect the much larger total number of particles simulated in this work is indispensible. Figure \ref{fig:khist} shows the distribution of entanglement points (i.e. the location of the kinks determined by the Z1 algorithm) as a function of distance in the $z$-direction from the center of the film. Two points are of immediate interest. Qualitatively consistent with the expectation that there are fewer chains to entangle with, near the film boundaries there is indeed a marked reduction in the entanglement density. Strikingly, though, there is \emph{not} a monotonic increase in entanglement density towards the bulk; all three films plotted have a maximum in the entanglement density approximately $4\sigma$ from the edge of the film (defining the edge of the film by the fit value of $\xi$). That distance is \emph{less} than the characteristic distance between kink points in the bulk, i.e., is less than the characteristic step size of the PP network.  From the location of that maximum to the center of the film there is a decrease in entanglement density that is qualitatively consistent with the theoretical predictions of Eq. \ref{eq:theory} as applied to geometrically confined films \cite{Sussman2014}. Below we will explore a more quantitative comparison, but note that a qualitatively similar result is obtained by considering not the density of entanglement points but the spatially resolved distance between successive entanglement points along a chain, $L_e(z)$. 

To attempt to connect these measurements with the theoretical model described above, we have analyzed the distributions of angular orientations of both the chain end-to-end vector and the individual PP segments. Key assumptions of the theoretical model include (a) that the orientational correlations at the end-to-end vector scale created by geometric confinement are directly communicated to the PP scale, and (b) that these orientational correlations at the PP scale alone allow one to predict changes in the properties of the entanglement network. We will see that despite how well the theory predicts global entanglement properties, neither of these assumptions is ultimately supported by the entanglement data generated by the Z1 algorithm.

\begin{figure}
\centerline{
\includegraphics[width=3.1in]{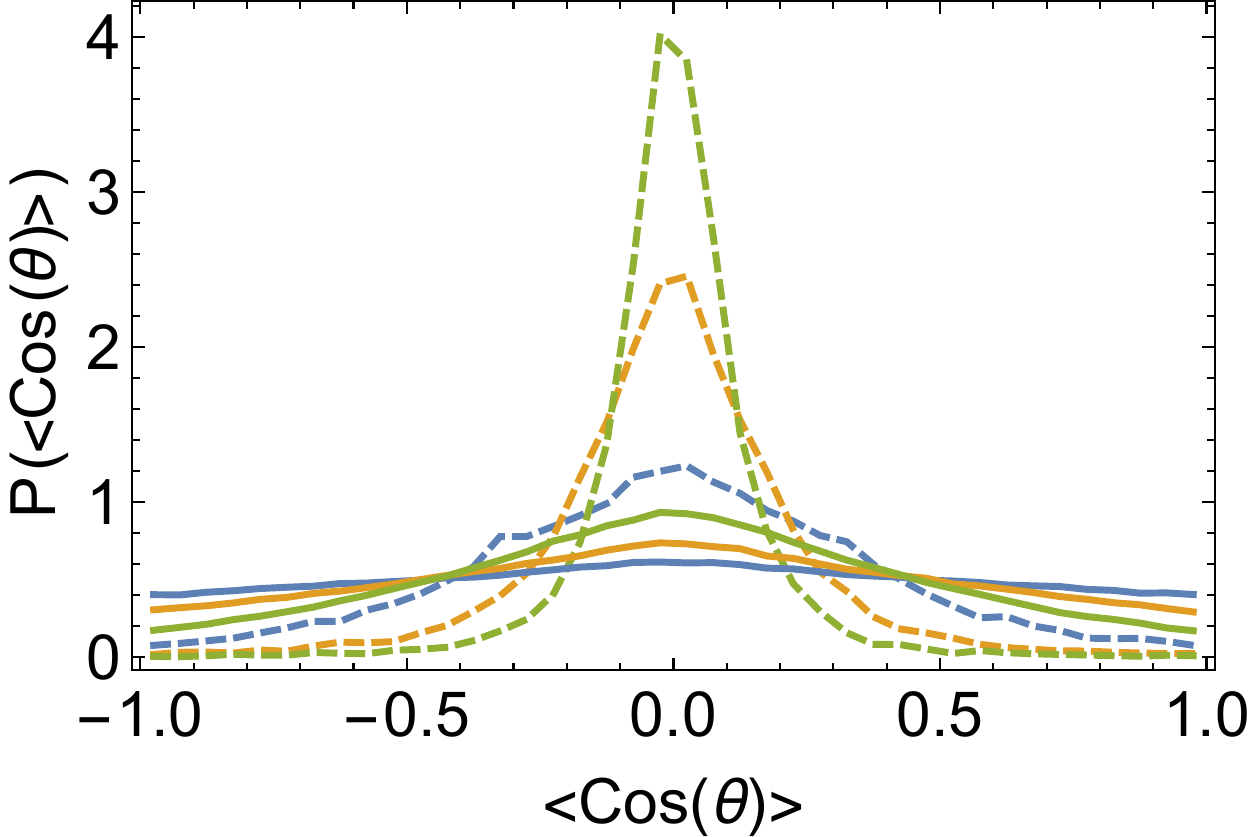}
}
\caption{\label{fig:allangle} Film-averaged distributions of the angle with respect to the $z$-axis of the chain end-to-end vectors (dashed curves) and PP segments (solid curves). From bottom to top each set of curves corresponds to $N=2000$ chains in free-standing films with effective thickness $h_{eff}=35\sigma$, $h_{eff}=20\sigma$, and $h_{eff}=13.3\sigma$.}
\end{figure}

Figure \ref{fig:allangle} shows film-averaged data for the probability distributions of chain and PP orientations with respect to the $z$ axis. It is immediately clear that the orientations present on the longest length scales are \emph{not} communicated to the PP length scale, at least for the primitive paths that are determined by the Z1 algorithm. This caveat is non-trivial, as the PP backbone detected by the Z1 algorithm may, in general, be different from the primitive path steps detected by other algorithms. In particular, the way the chain shrinking algorithms affect the geometry of the melt may be crucial in the orientational statistics of the derived primitive paths, and it is entirely plausible that this change is in the direction of a more isotropic distribution of angles. Nevertheless, at least by this metric, the chains are \emph{much} more strongly orientationally ordered at the end-to-end scale than they are at the primitive path scale. We note, though, that using the empirically observed distributions of chain-scale orientational order leads to a more accurate prediction of the film-averaged entanglement density than using the simplified orientational distributions assumed by Ref. \cite{Sussman2014} and reported as the curve in Fig. \ref{fig:Ztot}: using the observed angular distributions in Eq: \ref{eq:theory} leads to predictions of $\langle Z \rangle / \langle Z \rangle_{bulk} = 0.90,$ $0.79,$ and $0.76$ for the three thinnest $N=2000$ films, for instance, compared with the measured values of  $\langle Z \rangle / \langle Z \rangle_{bulk} = 0.92,$ $0.85,$ and $0.77$. In the context of the theory the success of the chain-scale predictions may be seen as a prediction of the ``super-coarse-grained'' needle mapping getting at some essential physics (albeit ignoring changes in the magnitude of the end-to-end vector of the chains under confinement, a severe approximation in the theory) \cite{Sussman2012}, or it may be a reflection that the chain end-to-end orientation is a better representation of the entanglement-scale orientational distribution than those reported by the Z1 algorithm.

\begin{figure}
\centerline{
\includegraphics[width=.3\linewidth]{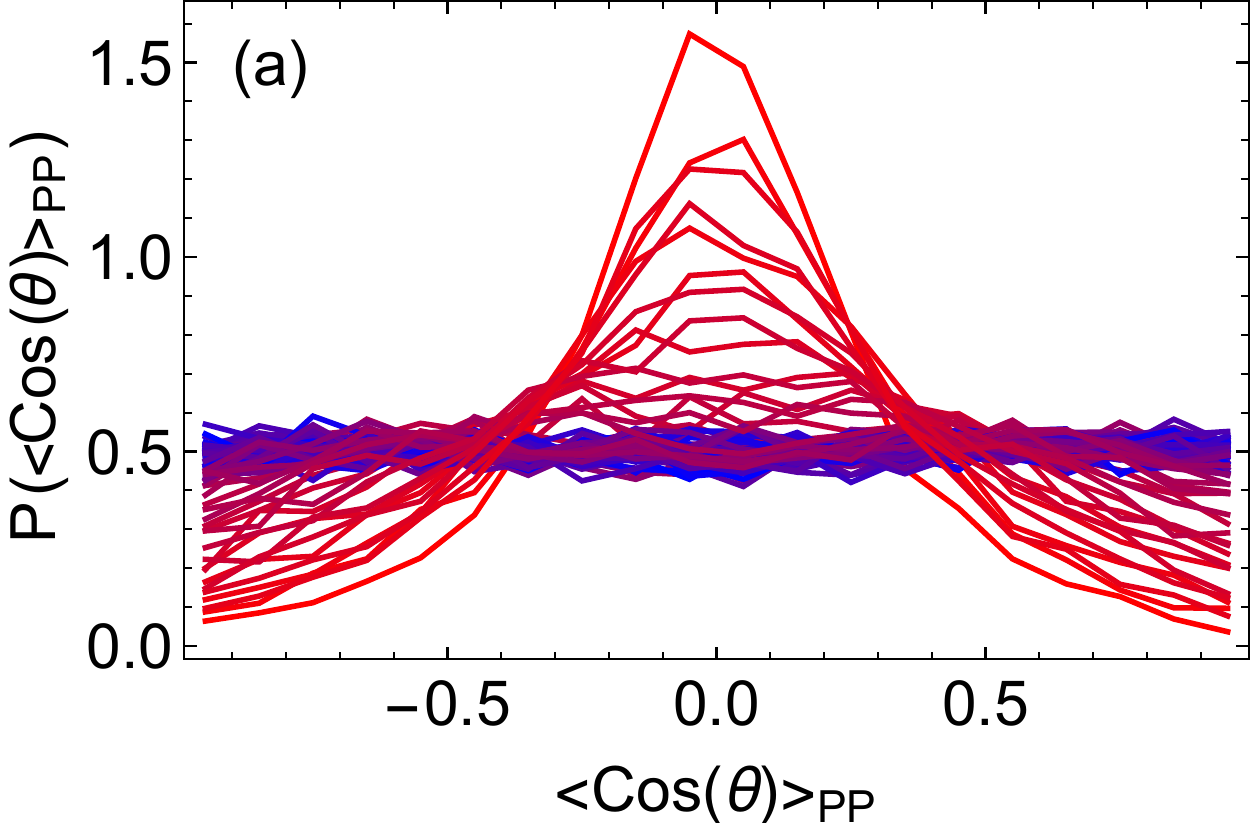}
\includegraphics[width=.3\linewidth]{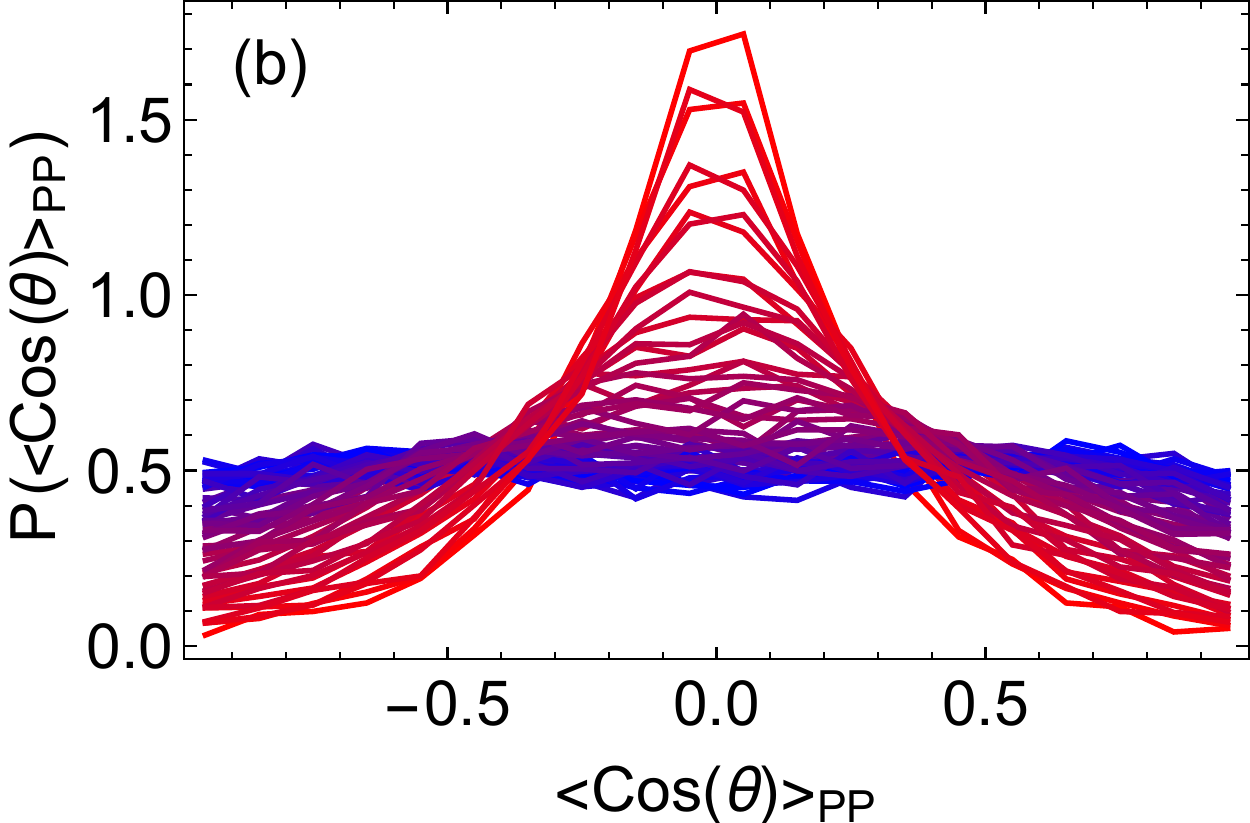}
\includegraphics[width=.3\linewidth]{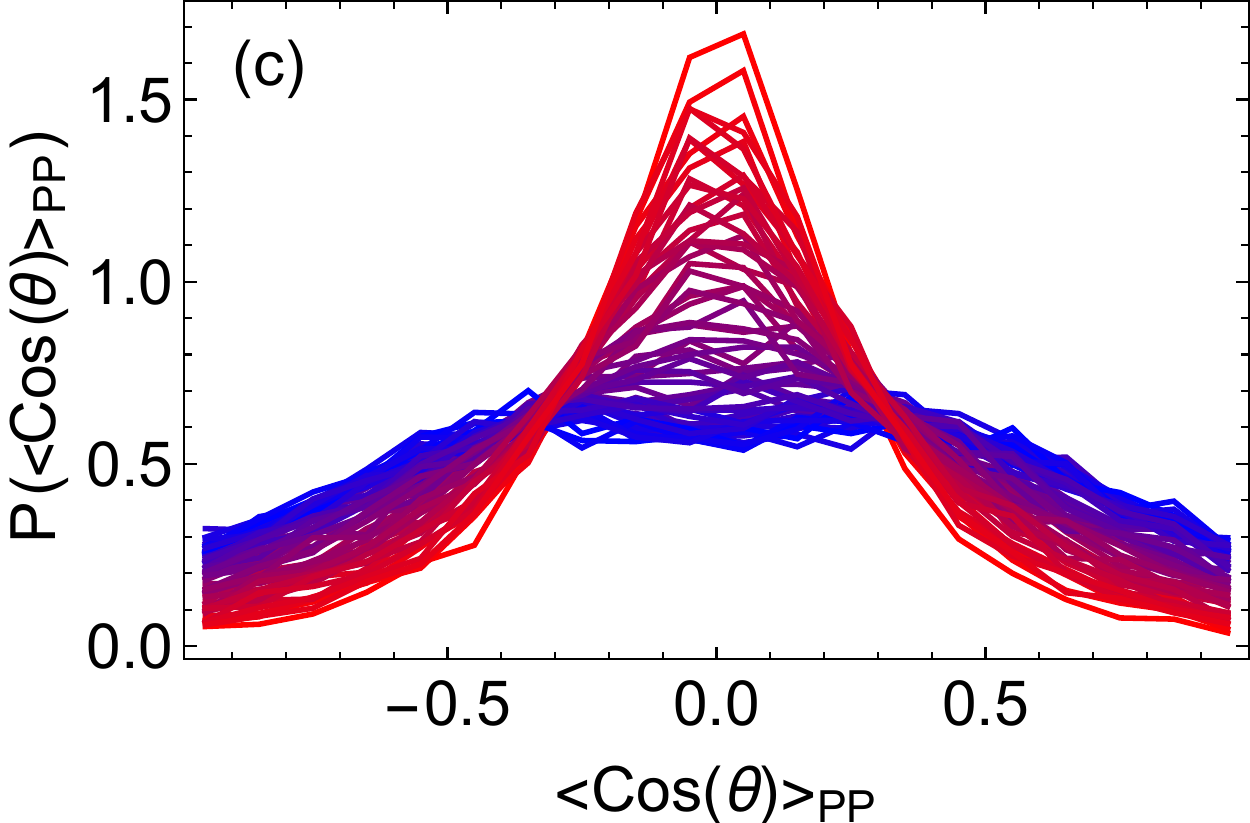}
}
\centerline{
\includegraphics[width=.3\linewidth]{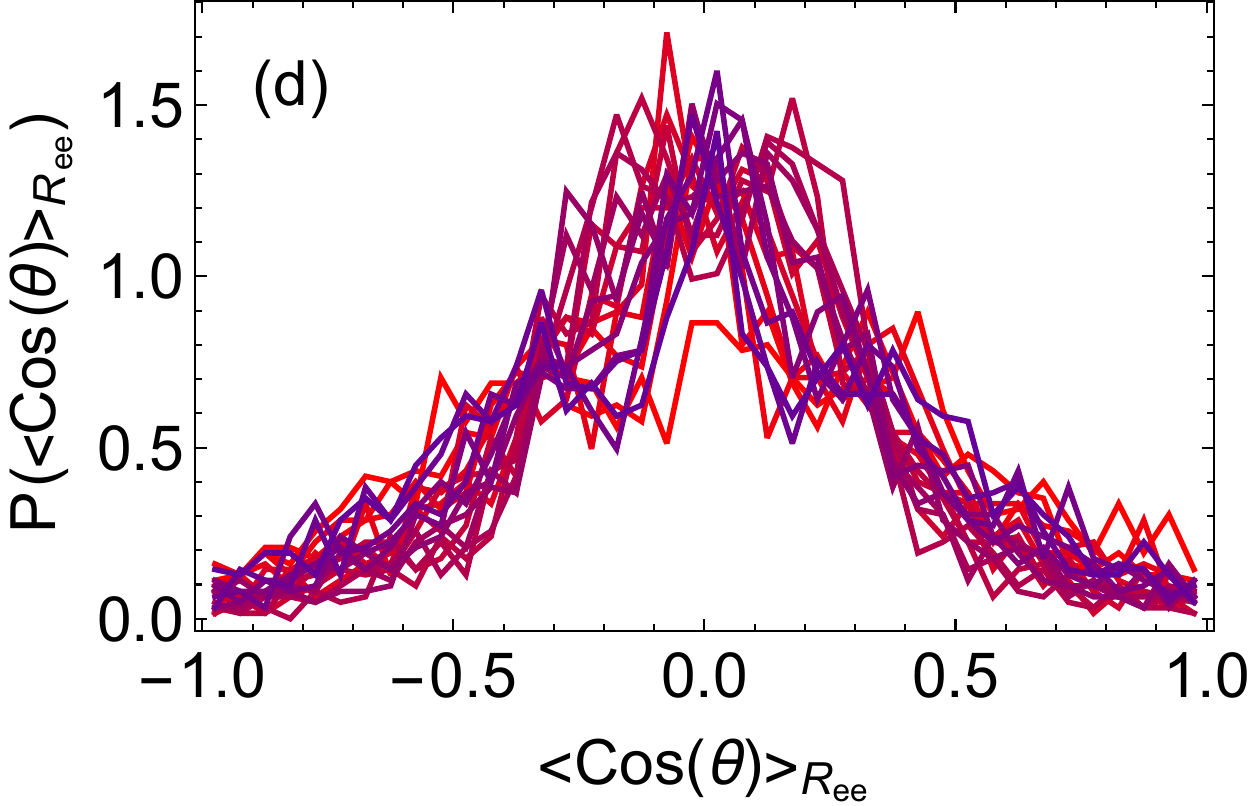}
\includegraphics[width=.3\linewidth]{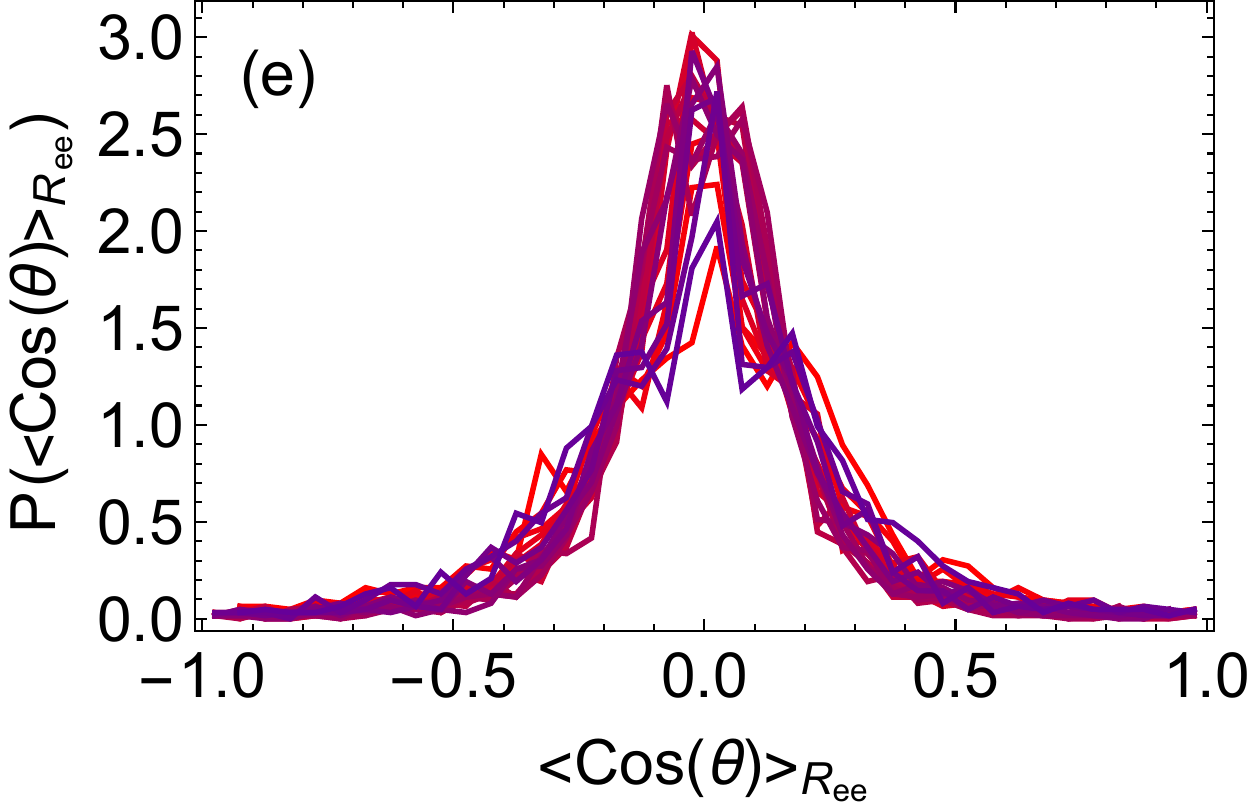}
\includegraphics[width=.3\linewidth]{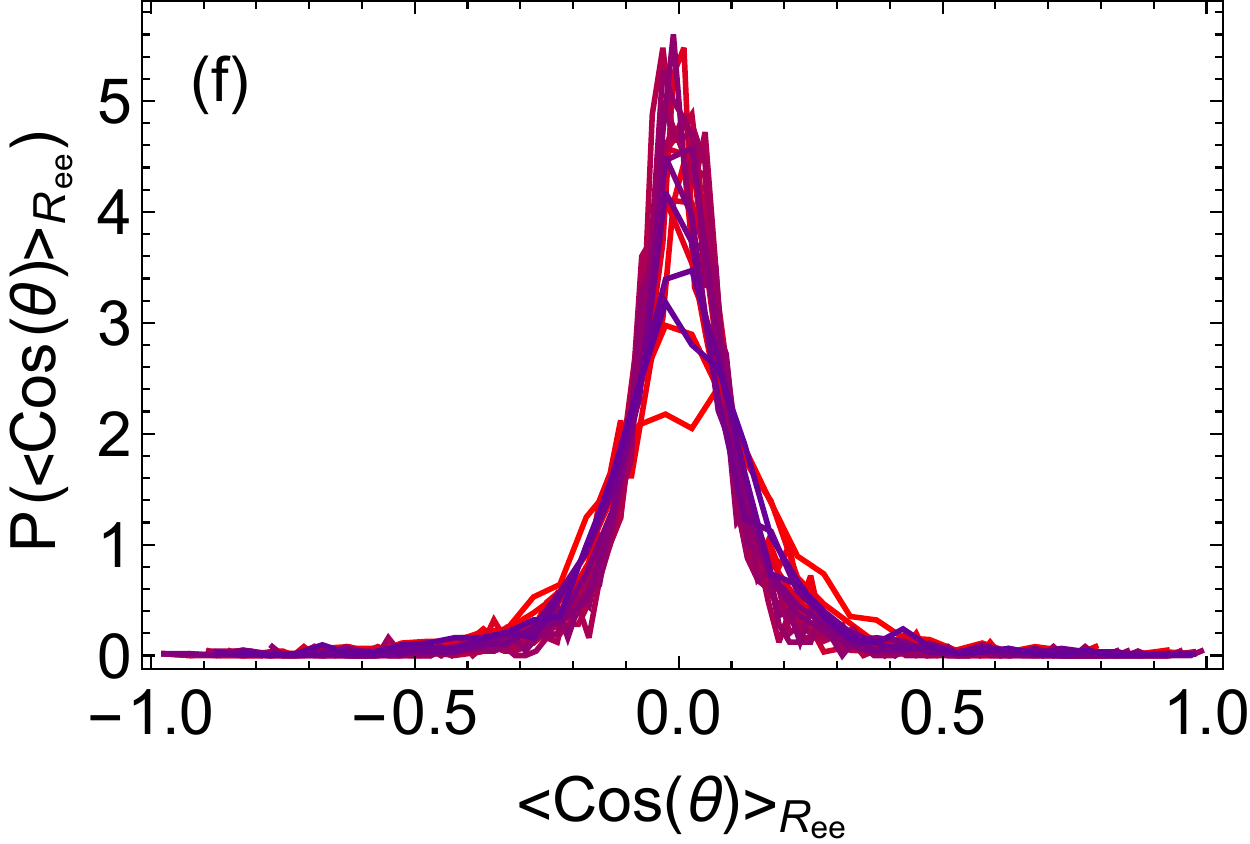}
}
\caption{\label{fig:angular} Layer-resolved distributions of the angle with respect to the $z$-axis of PP segments (a-c) and the chain end-to-end vectors (d-f). Each set of three plots corresponds to data from $N=2000$ chains in free-standing films with effective thickness $h_{eff}=35\sigma$, $h_{eff}=20\sigma$, and $h_{eff}=13.3\sigma$. Within each graph, redder (more peaked) curves correspond to layers closer to the edge of the film, and bluer (flatter) curves correspond to layers near the film center of mass.}
\end{figure}

These angular distributions can be spatially resolved into angular distributions as a function of the location of the end points of either the end-to-end vectors or the PP segments. This is shown in Fig. \ref{fig:angular}, where the data for each film is resolved into orientational distributions by layer in the film. The chain end-to-end vector orientational distributions, while noisy due to their more limited statistics, are slightly more sharply peaked around zero close to the center of the film, as expected by chain-scale theories of conformations adopted in confined geometries \cite{Cavallo2005b,Sussman2014}. In contrast, the PP-scale orientational distributions are much more strongly peaked at the film \emph{edges}, and are essentially flat in the center of the film. While perhaps surprising that there is a qualitative disagreement here, one might expect the chain-shrinking aspects of the Z1 algorithm to most strongly affect the angular distribution of the PP network near the edges of the film. The length scale over which the primitive paths adopt an essentially isotropic orientational correlations is $\sim 6\sigma$. This is slightly longer than the distance from the film edge of the entanglement density non-monotonicity shown in Fig. \ref{fig:khist}, and is shorter than $L_e$. 

\begin{figure}
\centerline{
\includegraphics[width=3.1in]{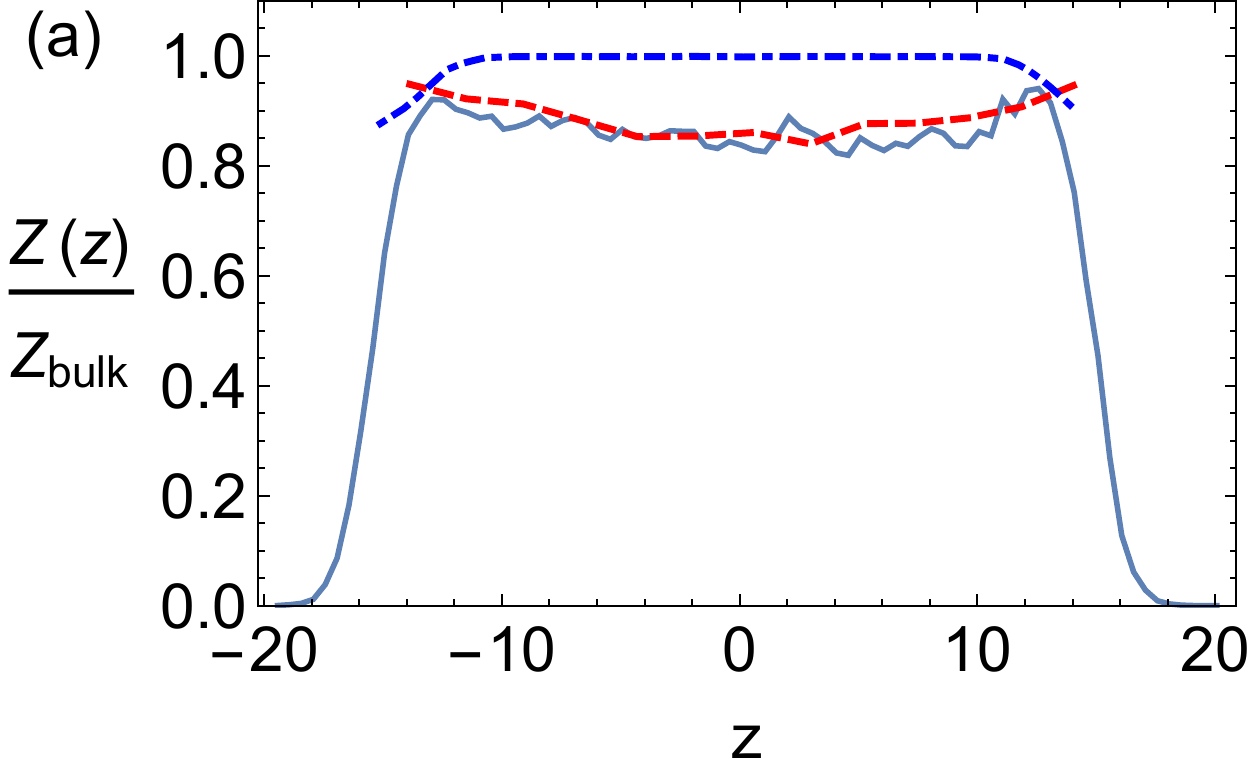}}
\centerline{
\includegraphics[width=3.1in]{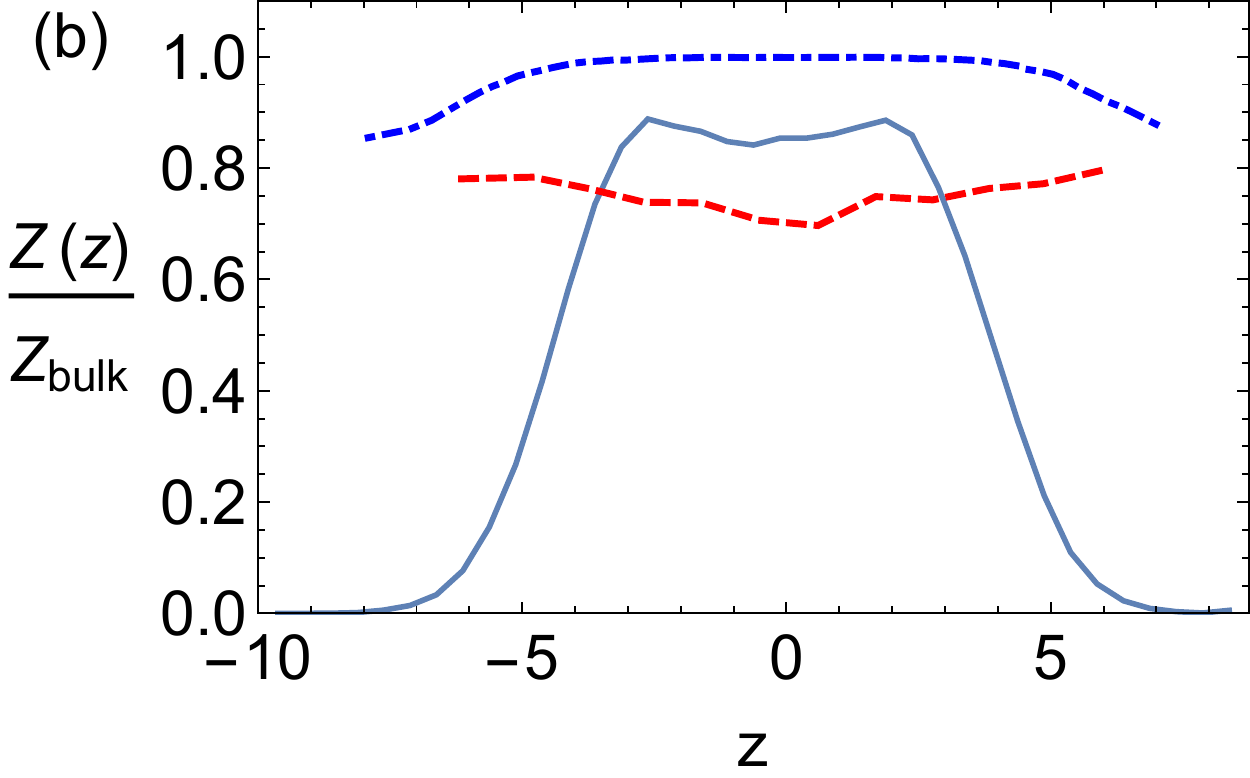}
}
\caption{\label{fig:theory} Layer resolved, normalized distance between entanglement points (solid curve) for the $N=2000$, $h_{eff} = 35$ film (a) and the $h_{eff}=13.3$ film (b). The red dashed curve is the layer-resolved prediction of Eq. \ref{eq:theory} using the chain end-to-end orientational distributions, and the blue dash-dotted curve is the layer-resolved prediction of Eq. \ref{eq:theory} using the PP end-to-end orientational distributions.}
\end{figure}

In principle the information in Fig. \ref{fig:angular} is the direct ingredient in making spatially resolved predictions of the entanglement statistics with the theory in Eq. \ref{eq:theory}: one simply imagines taking the integral over the angular distributions using the orientational distributions $g(\vec{u}_i)$ over whatever the local distribution is at that position in the film, rather than using the average over the entire ensemble. Figure \ref{fig:theory} carries out this analysis, averaging over the layer-resolved orientational distribution of both the primitive paths and the chain end-to-end vectors and comparing the normalized number of entanglement points in each layer with the theoryetical expectation for the $h_{eff} = 35\sigma$ and the $h_{eff} = 13.3\sigma$ film (whose behaviors are representative of the films studied in this work). Other than the precipitous loss of entanglement near the film edge, the chain-scale theory predicts the $h_{eff} = 35\sigma$ film properties very well. In contrast the chain-scale theory systematically over-predicts entanglement loss in the center of the thinner film and underpredicts it near the boundaries. The PP-scale theory is a qualitatively poor description of all films studied. The failure of the chain-scale theoretical predictions are most acute in regimes where PP-scale information is surely important, as indicated by the bulk properties discussed in the previous subseciton. The failure of the PP scale theory hinges on the connection between the geometry of the PP network revealed by the chain-shrinking algorithm and the true angular statistics of the entanglement strands in the melt.

\section{Discussion}\label{sec:disc}

In summary, we have studied both system-averaged and layer-resolved measures of entanglement statistics in a coarse-grained model of free-standing thin films made out of long linear polymers. The simulations show that, in accord with films confined by hard, amorphous walls, there is a pronounced decrease in the average number of entanglements per chain with decreasing film thickness. The film-averaged values of the entanglement density are reasonably well described by an existing theory \cite{Sussman2014}, although this requires taking chain-scale orientational distributions rather than the PP-scale orientational distributions that are more natural in the context of the theory. Further, we find that  existing proposals that collapse all entanglement results by plotting the entanglement density versus  $h_{eff}/R_{ee}$  are insufficient for the ultra-thin films simulated, and that an additional dependence on $h_{eff}/L_e$ must be taken into account. While the separable-variable hypothesis in this paper, $\langle Z \rangle / \langle Z\rangle_{bulk} = F_1(h_{eff}/R_{ee}) F_2(h_{eff}/L_{e})$, describes the data obtained in this paper fairly well, formulating a theoretical description of the combined dependence of the entanglement density on both length scales is an outstanding challenge.

While the film-averaged entanglement dilution is well-captured by the chain-scale orientational theory, the spatially resolved nature of this entanglement dilution is less well described. The non-monotonicity of the entanglement density versus distance from the center of the film observed in Fig. \ref{fig:khist} is a notable, qualitative \emph{prediction} of the theory \cite{Sussman2014}, but the quantitative details of the spatially-resolved nature of the entanglement dilution are more poorly captured in the theory. We then detailed the orientational statistics observed at both the chain- and PP-scale, noting that two of the key assumptions of the theory do not seem to be supported by the simulation data. We note that in these thin-film systems there is a weak segregation of chain ends near the interfacial layers of the film. While the chain ends themselves do not correspond to entanglements, this nevertheless leads to a slightly higher number of entanglements at a characteristic distance (the combination of the typical spacing between entanglements and the typical orientation of PP segments near the interface) from the free surface. However, this is a comparatively weak effect, affectly too small a percentage of the total entanglements in the system to account for the effects seen in Figs. \ref{fig:khist} and \ref{fig:theory}. 

A key observation of this work, and a strong candidate for the failure of the theory to correctly predict the spatially resolved properties studied in this work, is that both the film-averaged and the layer-resolved PP orientational distribution are very different from orientational distributions of the chain end-to-end vectors. It is qualitatively plausible that in Fig. \ref{fig:theory}B the correct theoretical curve lies in between the one derived from the chain end-to-end vector orientational distribution and the one derived from the primitive path information, in particular in the regime where the film thickness is not large compared to $L_e$. In polymer systems one can define a length-scale dependent orientational distributions, and this observation raises the possibility that it is not the orientational measures at the entanglement scale but some other length that dominates the problem.

A strong caveat to this, though, is that we have taken the primitive paths to be determined via the Z1 algorithm, and in particular the geometric properties of the primitive paths are likely to be strongly affected by any of the algorithms that use chain-shrinking methods to uncover the entanglement network. It is not surprising that the chain-shrinking algorithms tend to make the angular distributions of the primitive paths appear more isotropic than the entanglement strands in the unperturbed confined melt, but in this work we see that the algorithm may be completely erasing all informations about the geometry on this scale. Whether alternate approaches to finding the entanglement strands, for instance an alternate chain shrinking algorithm such as CReTA \cite{Creta}, or isoconifgurational time averaging of the chains \cite{Bisbee2011}, lead to strongly different geometric and orientational properties at the entanglement strand level will be an object of future study.

In general these results raise questions about the applicability of the theory, particularly when the primitive path orientational distribution does not match the chain-scale orientational distribution. We propose that new simulations of geometrically confined polymers, both free-standing and with hard walls, be conducted in order to further clarify these questions. We believe that simulations of reasonably long chains are crucial here, so that a clean distinction can be made between lengths on the order of the tube diameter and lengths of the order of the chain radius of gyration. Of particular interest would be a more complete surface describing the dependence on film-averaged entanglement statistics on both $h_{eff}/R_{ee}$ and $h_{eff}/L_e$. We also propose that new simulations and analyses be performed on entangled polymers under deformation; the entanglement dilution effect observed in steady-state continuous shear simulations \cite{Baig2010} is very well captured by the theory \cite{SussmanUnpub}, so in light of the present results a more critical evaluation of the theory, which would begin at the PP orientational scale, is called for. In closing, we reiterate that the connection between chain structure away from bulk, equilibrium conformations and either the statistics of the entanglement network or the dynamical properties of the system remains poorly understood (and not captured by either the classic Doi-Edwards tube model nor any recent extensions of it). We hope that the present results spurs the development of new theoretical descriptions elucidating these relationships.

\begin{acknowledgments}
This work was supported by the Advanced Materials Fellowship of the American Philosophical Society. The Tesla K40 used for this research was donated by the NVIDIA Corporation. The author would like to thank Rob Riggleman and Ken Schweizer for fruitful discussions and comments on this manuscript.
\end{acknowledgments}

\bibliography{filment_bib}

\end{document}